# Static Security Vulnerability Scanning of Proprietary and Open-Source Software: An Adaptable Process with Variants and Results

James J. Cusick
*Dept. of Electronic and Computer Engineering*
*Ritsumeikan University*
Kusatsu, Shiga, Japan
j.cusick@computer.org

*Abstract*—Software vulnerabilities remain a significant risk factor in achieving security objectives within software development organizations. This is especially true where either proprietary or open-source software (OSS) is included in the technological environment. In this paper an end-to-end process with supporting methods and tools is presented. This industry proven generic process allows for the custom instantiation, configuration, and execution of routinized code scanning for software vulnerabilities and their prioritized remediation. A select set of tools are described for this key DevSecOps function and placed into an iterative process. Examples of both industrial proprietary applications and open-source applications are provided including specific vulnerability instances and a discussion of their treatment. The benefits of each selected tool are considered, and alternative tools are also introduced. Application of this method in a comprehensive SDLC model is also reviewed along with prospective enhancements from automation and the application of advanced technologies including AI. Adoption of this method can be achieved with minimal adjustments and with maximum flexibility for results in reducing source code vulnerabilities, reducing supply chain risk, and improving the security profile of new or legacy solutions.

## I. INTRODUCTION

Providing secure software solutions remains a key challenge. While there are many design and operational approaches which aim to realize systems security [1][2] one of the most important of these relates to practices for software development which take into account the detection and remediation of software vulnerabilities [3][4]. This paper describes a part of this process in detail focusing on an industry proven and generic approach to conducting Static Application Security Testing (SAST) of large proprietary and open-source code bases.

Numerous studies have documented useful approaches to including SAST in development processes [4]. However, the method presented here, with demonstrated results, offers a practical industry-derived process template. After providing a generic SAST methodology, several project analysis experiences are documented including large scale open-source and proprietary code bases utilizing a specific set of tools. After this adaptable process is presented, examples are run with a set of specific tools.

The examples include the OSS projects Chromium and Gannon as well as a suite of proprietary industry applications. These samples range from about 1 KLOC to 6 million LOC. The results of the static vulnerability scanning of the source code from these examples include details of the identified code issues, their severities, and potential remediations. The methodology is documented in such a manner as to allow direct replication with identical steps and tools or to easily modify the process to meet local needs by substituting replacement tools or extending and/or automating the process steps to achieve nominally similar or better results as applied to a given team's own code base(s).

## II. BACKGROUND

The origins of the scanning practices outlined here began with the establishment of a holistic end-to-end security program developed at a large global information services company [5]. Aiming to standardize security practices and improve Cybersecurity preparedness and responsiveness, this program was modeled after an ISO 27001 control framework and included detailed policies and practices, governance, security staffing, staff education, and more. Furthermore, within this security program, a Secure SDLC (Software Development Lifecycle) was developed and implemented. This Secure SDLC was integrated into the existing Agile based software process already in use. Some key steps of this SDLC included [5]:

1. Security Requirements & Risk Analysis
2. Applying Secure Coding Best Practices
3. Conducting Threat Modeling
4. Static and Dynamic Vulnerability Scanning
5. Penetration Testing

Following the establishment of this process, and specifically the vulnerability scanning process, cyber threats have not stood still. Such attacks now include "heap sprays, double free, SQL injection, XSS, header smuggling" and more [6]. Nevertheless, the structure of the Secure SDLC we established remains both comprehensive and effective. There are some advances which can be considered on top of the core planks in our original Secure SDLC as mentioned by [6]:

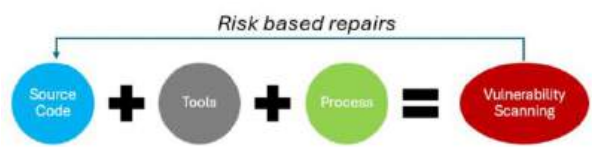

*Figure 1- A conceptual vulnerability scanning workflow*

- Safer languages and libraries.
- Improved sandboxes.
- Fuzzing practice.

However, the remainder of this paper focuses on the Static Vulnerability Scanning step. While the other steps are all equally important, static vulnerability analysis remains a significant area of payback in terms of development effort and risk reduction. Source code vulnerabilities also remain a major area of concern and security exposure for many systems as noted by [7][8]. The high-level conceptual process for this type of effort is shown in Figure 1.

The evolution of the Vulnerability Scanning approach began with initial research into methods, tools, and best practices. An experiment was first conducted using available freeware tools which led to the selection of Visual Code Grepper (VCG) as the demonstrator technology to launch the scanning initiative on several proprietary applications [9]. After conducting initial scanning and discussing results with the associated development teams, remediation plans were adopted. A business case was then developed for the purchase of more comprehensive commercially available products. As the Secure SDLC continued to mature the tooling became fully embedded into the overall SDLC and both improved the quality of results and the throughput of issue identification and resolution. The OSS examples below demonstrate the ability to repurpose the scanning method and its continued validity.

### III. Foundations of Static analysis for Security

Nearly forty years ago as a rookie programmer, the author was introduced to *lint* as a required tool in an SDLC (Software Development Lifecycle) used at Bell Labs. This utility was developed as a standalone program to detect code errors [10]. The program *lint* was later joined by similar tools and advances in compiler technology to routinely prevent coding errors. Development of static analysis tools advanced continuously since the introduction of *lint*. In fact, static analysis technology, algorithms, static analysis models, and subsequent tools have become markedly more powerful [11]. Peng states that static analysis tools use a wide variety of methods including: "lexical analysis, type inference, theorem proving, data flow analysis, model checking and symbolic execution" [11]. Today, static analysis is often an integrated capability within most IDEs, and a best practices-based code scan is a requirement prior to check-in.

Within the domain of Cybersecurity, as threat types have increased, the application of static analysis has become more critically important and more effective. In general there are several types of scanning or testing (static and dynamic) as well as active and passive scanning [12]. One category of scanning which is separate than static vulnerability scanning is Web vulnerability scanning. This type of scanning provides automated sensing of vulnerabilities in an active web or application environment [13] which is quite separate from the type of static vulnerability scanning focused on in this study.

Integrating vulnerability scanning in the development process has also become popularized within the current paradigm of the DevSecOps movement. Originally DevOps was a process notion that focused on the integration of development and operations [14]. This quickly led to the realization that security practices also needed to be blended into the approach, thus, the DevSecOps nomenclature was born [15]. Ideally, security considerations are now addressed at each step of the development and operations process supported by automation, tools, and continuous integration and delivery (CI/CD).

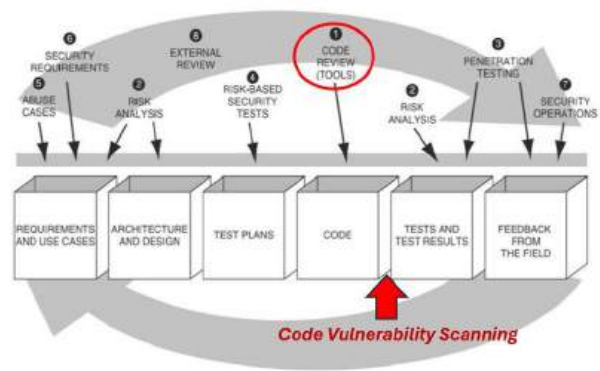

*Figure 2 – A Secure SDLC with an iterative Vulnerability Scanning stage highlighted [21].*

While integrating scanning into a development process does add value, such an approach is only part of a broader security strategy. In fact, most scanning tools produce a small ratio of serious vulnerabilities compared to all findings from a given pass [16]. Moreover, each scanning tool typically focuses on a set of anomalies or a pattern of errors [17]. Just like with anti-virus software these source code vulnerability scanning tools can only find issues they know about. Naturally, in the same manner that many fields are adopting LLMs (Large Language Models) and AI (Artificial Intelligence) to enhance their solutions, developers of static analysis tools are reporting improved accuracy in vulnerability detection by incorporating LLM based architectures [18]. Such advances may result in improved results for security.

Importantly, as Chess [17] points out: "A tool can also produce *false negatives* (the program contains bugs that the tool doesn't report) or *false positives* (the tool reports bugs that the program doesn't contain)." As a result, the developer must not only scan the codebase but review the output and address each vulnerability using a risk based analysis approach. In the remainder of this paper a templated methodology is presented using a sample toolset and several real-world examples.

### IV. Secure SDLC and Vulnerability Scanning

It has been said that for any software development team comprising more than three people, a process is required. Software Development Lifecycles define a process including inputs, outputs, handoffs, procedures, and more. Today's developers tend to favor Agile Methods [19] yet for any methodology, security has long been an important aspect of the development process. Viega [20] and others recognized this and helped mature the concept of the Secure SDLC [21]. A typical Secure SDLC is shown in Figure 2.

Naturally, each step of any SDLC is important and must be adhered to so as to produce software solutions meeting requirements. This is also true for a given Secure SDLC. Defining Security Requirements, Developing Risk-Based

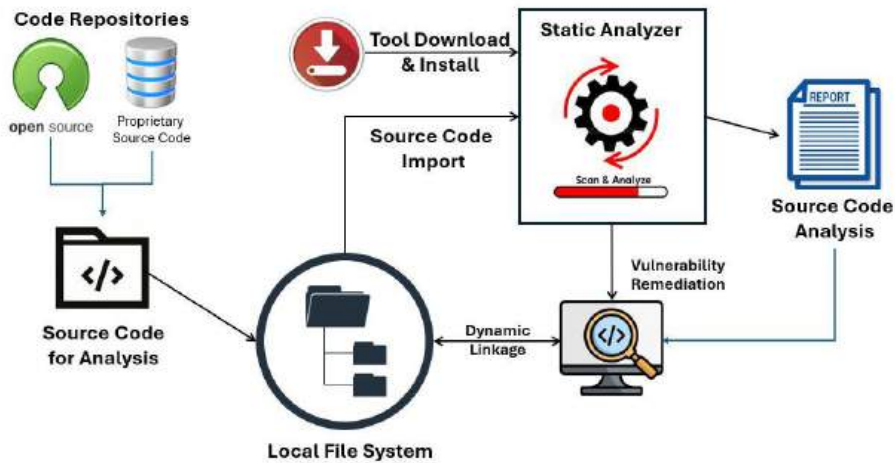

*Figure 3 - A Generic Process Model for Static Security Vulnerability Scanning of Proprietary and Open Source Software*

Security Tests, and conducting Penetration Tests are all vital to achieving secure software. However, this paper focuses on only one part of the Secure SDLC which is the Code Review piece, and particularly on vulnerability scanning.

## V. VULNERABILITY SCANNING AND REMEDIATION METHOD

Building on the capabilities of static analyzers, Secure SDLCs, scanning tools, vulnerability signature databases, and more, an end-to-end process for scanning source code can be established. This process is visualized at a high-level in Figure 3. The details of this process are provided in the steps below. Some variation around this process can be expected based on the needs of the development environment, development team, and current technologies, Nevertheless, this process with minor adjustments can apply to a wide range of Secure SDLCs where vulnerability scanning is required.

The process steps include:

1. **Select scanning toolset**: To select an appropriate tool for vulnerability scanning, first establish requirements around such needs as programming language support, platform, visualization capabilities, etc. Then, a review of available market offerings will provide a list of potential tools. Conducting a tradeoff analysis for the top tier candidates, including a hands-on trial is recommended [22].

2. **Install and configure scanner**: During an evaluation period the selected tool(s) may already have been installed and configured. However, in some cases further purchases, deployments, or licensing may be required. In most cases some learning is required on how to apply the scanning tool.

3. **Identify target codebase for scanning**: Oftentimes the target codebase has already been identified, especially in the case of proprietary software under development. The characteristics of this codebase may have helped drive the tool selection described in step one. However, this step may also be recurring or evolving based on the dynamic nature of the security analysis where the scope of code under consideration may change or grow given the version under development.

4. **Acquire codebase locally**: In nearly all cases the target codebase must be acquired (e.g., downloaded, fetched, or gotten) from a source code repository. For internally developed software this might be an SCM environment like CVS and for open-source software this might be an Internet based repository like Git. In any case, the code must be stored locally, preferably as a copy or branch for later manipulation and modification as part of the code ingestion and remediation phases upcoming.

5. **Import codebase into scanner for static analysis**: Once the target source code is available it can be imported into the static analyzer. This may require further configuration to adjust for source code location, programming language(s), etc.

6. **Conduct static analysis**: Typical analyzers provide numerous scanning options which can be applied at this point in the process. Scanning might cover all code, only non-comment source lines, or be filtered for specific languages. Depending on the size of the target code base this step can require some time to execute.

7. **Produce results**: Once static analysis is completed, results can be output. Tools may allow for a variety of output formats including inline results, CSV output, graphical representations, and more. Storing and versioning these results can provide additional value in analyzing progress over time.

8. **Review and analyze results**: Undoubtedly this step requires significant think-time and is arguably the most critical in providing for a secure solution aside from remediating the highest priority vulnerabilities discovered. A useful approach is to sort findings by priority or severity first. Then for each finding, determine the validity of the analyzer's result or warning. This can be done with a combination of experience in secure software development practices, common vulnerabilities, good coding practices, and current vulnerability inventories. In many cases, especially in large codebases, the individual conducting the scan may not be the one best able to understand the impact. Therefore, work needs to be divided and distributed. For many SDLC instances using an Agile method the results of the scan would be added to the Product Backlog for prioritization. Critically, some vulnerabilities arising from scanning might be of significant risk to the organization and would need to float to the top of any Backlog queue for immediate resolution.

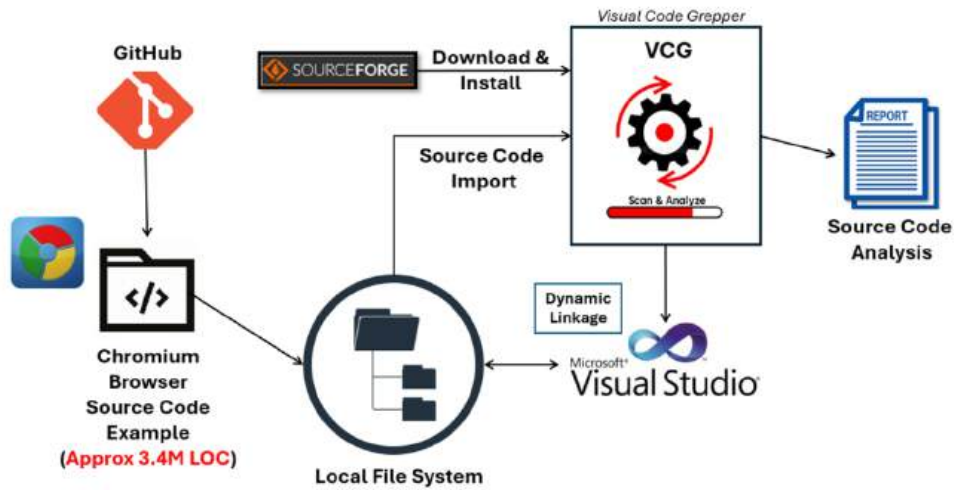

*Figure 4 – A Process Model Instance for SAST Scanning of Proprietary and Open Source Software*

9. **Interrogate source code flagged by static analysis**: Scanning tools are often integrated with code editors. This allows for navigation from vulnerabilities flagged by the static analyzer to the specific line of code generating the warning. Inspecting the source lines related to the warning can facilitate rapid remediation or, at minimum, speed up classification of the issue as a false positive or a valid issue.

10. **Remediate prioritized vulnerabilities**: Once the source of the vulnerability has been identified, reviewed, and potentially prioritized for a fix and assigned to a developer, a remediation can be applied. Such remediations can include noting that the vulnerability will not be addressed for a given reason, especially if the severity is low (i.e., a suspicious comment in the code). However, most vulnerabilities will need to be addressed via a code fix or sometimes a design change. In some cases, significant changes need to be applied and can require non-trivial effort to accomplish.

11. **Rebuild solution and test**: Critically, once the selected set of vulnerabilities has been remediated, the solution must be rebuilt and tested. Since any change to a code base might introduce defects (or reintroduce defects) the remediated code base must be verified.

12. **Rescan and repeat process as required**: Finally, this process must be repeated on every build. That is, as an integral part of the Secure SDLC whenever code is added, modified, or deleted, the entire code base must be re-scanned. This may not seem obvious at first, but even small code changes could re-instate a logic path which was considered unreachable earlier, and which might have a vulnerability. More typically, newly written code can introduce new vulnerabilities. And lastly, new vulnerabilities are discovered regularly, and updated vulnerability definition files may find new issues in previously clean code.

## VI. CODE ANALYSIS EXAMPLES WITH RESULTS

Given the above generic process template, a specific instance of a scanning environment with known bounded capabilities can be derived and implemented. This environment can then be applied to several practical codebase examples. In this section a specific implementation of a scanning environment is defined, and several examples of its usage are presented. Two separate OSS examples (one large and one small) are demonstrated as well as four independent proprietary applications. This realized instance of the generic process and these examples provide a concrete roadmap for implementing this environment model and/or how to vary the implementation based on local technological needs.

### A. A Vulnerability Scanning Tool Instance: VCG

Starting with the process model template of Figure 3, each generic component of the process now needs to be converted to a specific element. The first tool component is the repository. For the OSS examples this is conducted via the GitHub repository. In the case of the proprietary software example the source code resided in Microsoft's Team Foundation Server (currently Azure DevOps Server). Naturally, these source code repositories are replaceable with any code repository or multiple source code repositories simultaneously or in parallel.

The next significant tool instance within the process is the vulnerability scanner itself. For purposes of this research, VCG (Visual Code Scanner) was employed. This tool is available from SourceForge as freeware. VCG offers adequate features for essential scanning requirements and provides export of results in multiple formats as well as dynamic linkage to code editing tools. In this case, Microsoft VisualStudio was utilized for investigating vulnerabilities. The entire process, as implemented to support the first example of Chromium, is displayed in Figure 4.

### B. An OSS Vulnerability Scanning Example: Chromium

For our first example we analyze Chromium, an open-source software project developed primarily by Google. This code base underlies more than a dozen web browsers, some of which are highlighted in Figure 5 [23]. These browsers, led by Chrome, dominate global browser market share at 77.53%

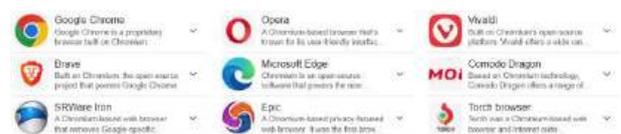

*Figure 5 – Primary Chromium-based browser implementations [23]*

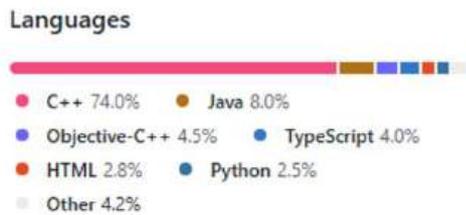

*Figure 6 - Proportion of language types found within the Chromium code base via VCG analysis*

[24]. Other popular browsers contributing to this level of usage include Edge, Vivaldi, and Opera. The Chromium project was selected for this analysis due to its broad recognition, importance, and scale.

To initiate this analysis, a version of the Chromium project [25] was downloaded on April 13th, 2025. The code base was then ingested into the VCG tool which indicated a total of 27,184 files loaded. Following this step a full scan of both comment and non-comment lines of code was initiated. This step required approximately 15 minutes to complete on a recent model laptop configured with an Intel i7 processor and 16 GB RAM. This analysis produced a variety of outputs including a breakdown of the programming languages found within the project and the proportionality of their usage as seen in Figure 6. Importantly, nearly 80% of the project is implemented in C++ or a related language with less than 10% written in Java. In terms of scale, the project totaled 5.9 million LOC (Lines of Code) of which 4.1 million were NCLOC (Non-Comment Lines of Code).

Most importantly, in addition to these metrics, the static analysis from VCG also provided a detailed listing of all vulnerabilities detected in the codebase. Overall a total of 1,460 potential issues were identified which means there was 1 potential vulnerability per 2,808 LOC (or 1:2808). This vulnerability density may vary from application to application. In this case it would make sense if implementing a Chromium based browser to certainly scan the project code and decide which vulnerabilities need attention.

Vitally, VCG assigns criticalities or severities to each vulnerability on a scale of seven from suspicious comment to critical. As can be seen in Figure 7, only four issues were categorized as Critical and 104 as High. This limits the exposure of this application yet each of these issues should be prioritized for review and possible remediation. Of note as well are the large number of "suspicious comments" detected. Upon review, many of these incidents may simply be instructions to future developers on where to hook in new functionality. In some cases they might include information that could harm the operations of the application although this would require one-by-one verification.

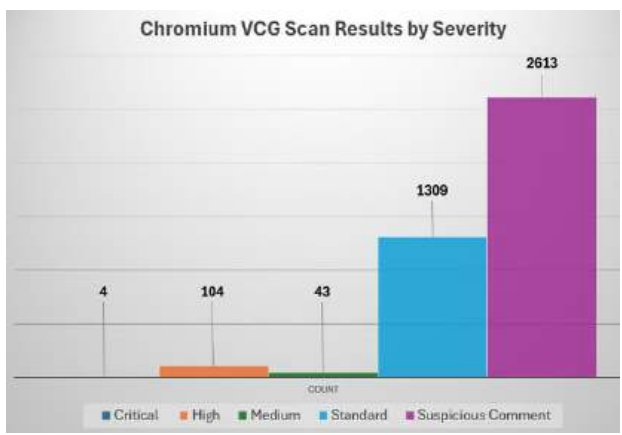

*Figure 7 – Chromium vulnarability totals by severity as per VCG analysis*

*Table 1 – Example of high severity vulnarabiltiy found within the Chromium codebase*

| Item | Description | Line | Code |
|---|---|---|---|
| **Unsafe Use of memcpy Allows Buffer Overflow** | The size limit is larger than the destination buffer, while the source is a char* and could allow a buffer overflow to take place. | 111 | std::memcpy(buffer, str, length); |

A sample Severity 1 vulnerability report is provided in Table 1. In this instance a *memcopy* call has the potential to introduce a buffer overflow which might be exploited. The best practice in this case would be to add precautionary code to check the size of both the buffer and the string to ensure alignment. Note that the tool in its full report (in addition to that found in the enclosed table) provides further information on each vulnerability including severity, file path, file name, etc., allowing for pinpoint analysis. Ideally, the assigned developer would investigate each high priority vulnerability and determine if it required remediation.

*C. A Second OSS Vulnerability Scanning Example: Genann*

As a second open-source software example, Genann was selected [26]. This neural network library represents a much smaller codebase of only 682 LOC. The same process was applied to the vulnerability scanning which included 1) downloading Genann from GitHub; 2) starting the VCG tool; 3) loading the Genann source code; 4) scanning the code; and 4) reviewing the results. In this case the scan was completed nearly instantaneously as the code sample was so small.

The scan uncovered a total of six potential vulnerabilities. These issues came with high severities and related to potential memory mismanagement. Cases of using *malloc* without any call to *free* as well as cases where multiple *frees* were executed on the same memory store were highlighted. On such a small code base this translates into 1 vulnerability per LOC (or 1:27). This is a much higher density than found within our first example of Chromium. Again, while some of these hits might not require remediation it certainly indicates that any given package may have relatively more or less vulnerability exposure. This should be considered when integrating code from public repositories.

*D. Vulnerability Scanning Example: Proprietary SaaS*

For a final scan example, a suite of proprietary applications is analyzed. As opposed to the first two examples from the open-source community these applications were developed by industry teams within a common corporate culture over many years. The codebases under review were developed for commercial purposes where the domain centered on business services utilizing a SaaS model. The applications initially applied a classical SDLC and later an

Table 2 - Application code totals and vulnerabilties per proprietary SaaS application reported by VCG

| Application | App SLOC (Source Lines of Code) | Potentially Dangerous Code Lines | Total Reported Potential Issues |
|---|---|---|---|
| App1 | 1,048,017 | 214 | 371 |
| App2 | 797,378 | 1,072 | 1,409 |
| App3 | 756,009 | 2,942 | 2,942 |
| Libraries | 298,673 | 128 | 254 |
| Total | 2,900,077 | 4,356 | 4,976 |

Agile methodology. At the time this analysis was initially conducted these applications were not utilizing any open-source components, but they did integrate some purchased application libraries or utilities. However, those were pre-compiled binaries and were not included as part of the vulnerability scanning exercise.

*1) Proprietary SaaS Scan Approach*

The source code scan of several web applications was conducted in order to gauge the scope of any vulnerabilities in a proprietary code base. The code was extracted from the then current production support branch of each of the three applications and from an internal utility catalog used by all three end user applications. The scan was carried out with Visual Code Grepper (VCG) just like the earlier open-source examples.

For the three main applications, along with their shared libraries, there was approximately 2.9 million LOC present. The three major applications ranged from 750,000 LOC to 1 million LOC as per Table 2. The primary languages comprising the code base were C# and SQL. Within this code base VCG reported 4,976 issues distributed fairly evenly across the applications with App 1 having the lowest presence of vulnerabilities as shown in Table 2. This translates on average to one security vulnerability for every 582.8 lines of code (or 1:582). Across this sample set, the vulnerability densities per line of code ranged from 1:2825 to 1:566 to 1:257 to 1:1176, indicating that even similar teams may produce different quality results.

*2) Proprietary SaaS Scan Findings*

Within the nearly 5,000 vulnerabilities uncovered by the scan, all were Medium severity issues or below. There were no Critical or High severity issues reported. However, some issues seemed to require attention in the short-term as they had a potential for compromise or exploitation. Additionally, some issues were straightforward to solve for or could be solved in a central location thereby allowing for rapid progress in reducing the risk profile with limited effort.

To provide a fuller sense for the type of vulnerabilities reported, see Table 3. In this table a range of vulnerabilities are listed along with severities and with their frequency of occurrence. That is, for each of the finding types there were numerous occurrences. For example, more than one apparent hardcoded password instance, numerous questionable comments, several Cross Site Scripting issues, and more.

Just as with the OSS examples, each of these potential vulnerabilities requires review, confirmation, and potential remediation. Each example seen in Table 3 introduces pros and cons to consider. In practice, the process of flagging these issues and discussing the relative merits of remediation versus accepting the risk is highly valuable in itself. For example, the risk exposure of running production code with .Net debugging enabled may be debated but there is no substantial reason to deploy in this configuration as it opens a new attack vector where none needs to be provided.

Table 3 - Sample vulnerabilties by severity from proprietary Saas scan with frequencies

| Priority | Vulnerability Type | Count |
|---|---|---|
| MEDIUM | Potentially Unsafe Code - .NET Debugging Enabled | 1 |
| MEDIUM | Potentially Unsafe Code - Potential XSS | 2 |
| LOW | Potentially Unsafe Code - Thread Locks - Possible Performance Impact | 7 |
| STANDARD | Potentially Unsafe Code - LoadXml | 8 |
| STANDARD | Potentially Unsafe Code - Potential TOCTOU (Time Of Check, Time Of Use) Vulnerability | 8 |
| STANDARD | Potentially Unsafe Code - URL Request Gets Path from Variable | 1 |
| SUSPICIOUS COMMENT | Comment Indicates Potentially Unfinished Code | 63 |

Table 4 contains a few sample detailed messages for specific vulnerabilities detected in the scan. These samples provide a little further context around what findings from the scan look like. They include the severity, a title for the warning/error, a brief description, and the specific line of code in question. Within the actual tool environment the path, file name, and line number is also provided.

*3) Impact of Scan Process for Proprietary Apps*

Upon introduction of this scanning process to the development of these proprietary applications the local SDLC itself was modified to incorporate ongoing scans. The original scan documented here was used as a baseline for the rate of vulnerabilities encountered. Rapidly, the freeware scanning tool VCG was replaced by commercial tools including IBM

Table 4 – Sample vulnerabiltiy reports from VCG on proprietary SaaS application

| SEVERITY: Vulnerability (description; code snippet) |
|---|
| 1. **MEDIUM: Potentially Unsafe Code - Appears to Contain Hard-Coded Password** <br> The code may contain a hard-coded password which an attacker could obtain from the source or by dis-assembling the executable: <br> *public const string R1ResetPassword = "MYPASSWORD";* |
| 2. **MEDIUM: Potentially Unsafe Code - Unsafe Password Management** <br> The application appears to handle passwords in a case-insensitive manner. This can greatly increase the likelihood of successful brute-force and/or dictionary attacks. <br> *string strUserPassword = txtCurrentPassword.Text.ToUpper();* |
| 3. **MEDIUM: Potentially Unsafe Code - Insecure Storage of Sensitive Information** <br> The application used standard strings and byte arrays to store sensitive transient data such as passwords and cryptographic private keys instead of more secure: <br> *SecureString class.String key = null;* |

AppScan (now HCL AppScan) and SonarCube, further demonstrating the flexibility of the plug-and-play nature of the process and providing for refined vulnerability discovery.

Developers were also educated on how to execute their own scans to prevent known vulnerabilities from being released into production. Additionally, steps were taken to actively prevent new vulnerabilities from being introduced by developing an integrated Secure SDLC as mentioned above. Aside from conducting routine vulnerability scans upon check-in of code, secure coding practices were introduced such as reviewing OWASP lists, regular Dynamic Application Security Testing (DAST), and improved security testing.

The original 1:582 ratio of vulnerabilities to source lines also served as a useful reference metric for planning and effort estimation. In general, such measures can be used to gauge potential vulnerability densities and the concomitant remediation work required in related applications. Naturally, this vulnerability density rate changes over time. However, since additional proprietary applications were built in a consistent manner by similar teams and processes, this same ratio of vulnerabilities to source code can also be applied to for planning and secure development estimation purposes.

## VII. Discussion

The generic and modifiable methodology presented here allows for effective discovery of source code based security vulnerabilities. This methodology also allows for the scanning and analysis of both open-source software and proprietary code bases. Furthermore, while the method does provide an end-to-end process which is partially automated, additional steps within the process could be automated and, importantly, each of the technologies deployed and demonstrated can be substituted based on technology evolution, user conditions and implementation requirements.

As currently described and demonstrated, the methodology provides an effective means of implementing a key function of the DevSecOps lifecycle. In specific, the snapshot of a code base, its static analysis, determination of indicated vulnerabilities, and a dynamic linking to implicated source lines via an IDE to enable remediation are all provided. This process can be enhanced by existing technologies such as Cloud based computing at scale [27], the application of Machine Learning (ML) to more rapidly detect and remediate issues [28], end-to-end automation attempts, and application of LLMs and related AI technology for improved vulnerability analysis and corrections [29][30].

An important note from the examples worked using this process, especially from the open-source community, is the impact on secure supply chains [31]. The results of these examples yielded numerous potential vulnerabilities throughout the freely available code bases selected. This puts the onus on the developers integrating these code bases to carefully screen for security vulnerabilities and not to blindly incorporate unscanned code into their custom solutions.

## VIII. Conclusions

This study introduced a methodology to conduct source code vulnerability analysis using a generalized process and supporting interchangeable tooling components. The methodology covers source code acquisition, static analysis, vulnerability review, and remediation. The process was first presented in a generic form, and then provided using a specific set of tools and technologies. Three separate examples were demonstrated with this methodology and specific toolset covering both OSS and proprietary applications.

Results from each process iteration were discussed including the specific types of vulnerabilities found, vulnerability densities, and potential remediation paths. This process can benefit development teams looking to implement a Secure SDLC which can support a DevSecOps vision by efficiently and flexibly adapting this method. Finally, Additional research and experimentation in applying advanced tools and automation techniques (such as with scripting or ML) on top of this methodology are conceivable and can further improve the timeliness and completeness of scanning results and reduce effort. Nevertheless, this generic method can provide a base pattern for instituting a scalable and modifiable vulnerability detection and remediation approach today and for the future.